# Second generation Robo-AO instruments and systems

Christoph Baranec*[a], Reed Riddle[b], Nicholas M. Law[c], Mark R. Chun[a], Jessica R. Lu[a], Michael S. Connelley[a], Donald Hall[a], Dani Atkinson[a], & Shane Jacobson[a]

[a]Institute for Astronomy, University of Hawai'i at Mānoa, Hilo, Hawai'i 96720-2700 USA; [b]Caltech Optical Observatories, California Institute of Technology, Pasadena, California, 91125 USA; [c]Department of Physics and Astronomy, University of North Carolina at Chapel Hill, Chapel Hill, NC 27599-3255, USA.

## ABSTRACT

The prototype Robo-AO system at the Palomar Observatory 1.5-m telescope is the world's first fully automated laser adaptive optics instrument. Scientific operations commenced in June 2012 and more than 12,000 observations have since been performed at the ~0.12" visible-light diffraction limit. Two new infrared cameras providing high-speed tip-tilt sensing and a 2' field-of-view will be integrated in 2014. In addition to a Robo-AO clone for the 2-m IGO and the natural guide star variant KAPAO at the 1-m Table Mountain telescope, a second generation of facility-class Robo-AO systems are in development for the 2.2-m University of Hawai'i and 3-m IRTF telescopes which will provide higher Strehl ratios, sharper imaging, ~0.07", and correction to $\lambda$ = 400 nm.

## 1. INTRODUCTION

The prototype Robo-AO system at the Palomar Observatory 1.5-m telescope is the first fully automated laser adaptive optics instrument[1]. It commenced scientific operation in June 2012 and has since executed more than 12,000 observations at the ~0.12" visible-light diffraction limit[2-5]. Among these observations are the largest diffraction-limited surveys of stellar multiplicity in the local solar neighborhood[1], surveys of nearby companions to *Kepler* exoplanet host candidates[6] and several other commissioning science programs[7-11]. The Palomar system will be augmented with two new infrared cameras during the summer of 2014: a 2 arcminute field-of-view camera using a Teledyne HAWAII-2RG detector sensitive to wavelengths from 1 to 2.5 microns; and a very-low noise, high-speed Selex/SAPHIRA electron avalanche photodiode (eAPD) detector[12,13] to be used for infrared tip-tilt sensing. Both cameras will widen the spectral bandwidth of observations and enable deeper visible-light imaging using adaptive-optics-sharpened infrared tip-tilt guide sources, necessary for Robo-AO to complete the survey of all *Kepler* exoplanet host candidates by the fall of 2014. Additionally, a clone of the prototype Robo-AO system is being built for the IUCAA Girawali Observatory 2-m telescope[14], and a natural-guide star only variant, KAPAO, is deployed at the Pomona College 1-m Table Mountain telescope[15,16].

Two new second generation facility-class Robo-AO systems are currently in development, one for the University of Hawai'i 2.2-m telescope and another for NASA's 3-m IRTF. The new systems will build upon lessons learned from the prototype system, use identical improved components (e.g. low-noise gated CCDs, larger actuator count deformable mirrors, higher efficiency optical coatings) to mitigate development costs and exploit the extraordinary seeing conditions that exist at the summit of Manuakea. They will routinely provide improved Strehl ratios, deeper contrasts, sharper imaging (~0.07"), and short visible wavelength correction to 400 nm, all while extending the typical tip-tilt guide source limit by 1 to 2 magnitudes over the prototype. In addition to continuing ongoing massive scientific surveys, the second generation Robo-AO systems will be used for high-cadence monitoring of weather and volcanism within the solar-system, characterization of asteroids and supernovae discovered by the Palomar Transient Factory[17], Zwicky Transient Facility[18], Pan-STARRS[19] and The Asteroid Terrestrial-impact Last Alert System (ATLAS)[20], and for validating the tens of thousands of exoplanet host candidates that will be discovered by the *Kepler* K2 mission[21] and future Transiting Exoplanet Survey Satellite[22].

*baranec@hawaii.edu; phone 1 808 932-2318; http://high-res.org; http://robo-ao.org

## 2. ENHANCED ROBO-AO CAPABILITIES

The combination of an invisible, pilot-safe ultraviolet laser guide star adaptive optics (AO) system and integrated low-noise, high-speed visible and infrared imaging arrays (see table 1), which double as tip-tilt sensors, allows for a wide range of high-resolution imaging capability. While the prototype Robo-AO had an electron-multiplying CCD (EMCCD) camera and an external infrared port, the new Robo-AO systems will relay a corrected F/41 beam which will be split by a visible/infrared dichroic (3 arc minute diameter for the UH 2.2-m and 1 arc minute for the IRTF). The visible light path will normally be imaged by the internal EMCCD camera, but can also be directed out of the instrument to an external visible instrument port. Similarly, the infrared light path will be imaged onto the internal HgCdTe Selex Avalanche PHotodiode InfraRed Array (SAPHIRA)[13,23], or directed to an external infrared instrument port.

Table 1. Format and characteristics of internal imaging and tip-tilt detectors within the new Robo-AO systems.

| Detector | Format | Field | PixelScale | ReadNoise | Full frame rate | Tip-tilt rate | Initial filters |
|---|---|---|---|---|---|---|---|
| UH 2.2-m Robo-AO | | | | | | | |
| EMCCD | $1024^2$ | 31"x31" | 0.030" | <1e- | up to 25.6 Hz | to 400 Hz | g, r, i, z, Hα, SII, r+i+z |
| SAPHIRA | $256^2$ | 14"x14" | 0.055" | 1-3e- | up to 100 Hz | to 8 kHz | Y, J, H, FeII, open |
| 3-m IRTF Robo-AO | | | | | | | |
| EMCCD | $1024^2$ | 23"x23" | 0.022" | <1e- | up to 25.6 Hz | to 400 Hz | r, i, z, Hα, SII, r+i+z |
| SAPHIRA | $256^2$ | 10"x10" | 0.040" | 1-3e- | up to 100 Hz | to 8 kHz | Y, J, H, FeII, K, open |

The new Robo-AO systems will offer several distinct observation modes, targeting different areas of wavelength and sky coverage parameter space, distinguished by use of either an on-axis science target or a nearby field star for tip-tilt wavefront sensing. Our best imaging performance will be obtained for targets that are brighter than $m_V$=14 (MV): we can actively correct image motion using tip-tilt measurements from one of the two cameras or we can read out both detectors at 25.6 Hz and use post-facto shift-and-add image processing algorithms to mitigate residual image motion errors. For targets as faint as $m_V$=18 (MV), post-facto methods will work with broad filters (e.g. r+i+z in the visible), or active tip-tilt correction by one of the cameras will be required to maintain close to diffraction-limited resolution in standard filters; the latter method will be used for imaging very faint objects with a nearby tip-tilt guide star. Figure 1 shows typical visible-light diffraction-limited images captured with the prototype Robo-AO and figure 2 shows the achieved 5-σ contrast ratio vs. separation in the search for faint companions to stars. The UH 2.2-m Robo-AO system will enable the detection of companions at closer separations, ~0.07" vs. 0.12-0.15" in the visible, and achieve deeper contrasts, e.g. an extra ~1.8 magnitudes at 0.5 arc seconds.

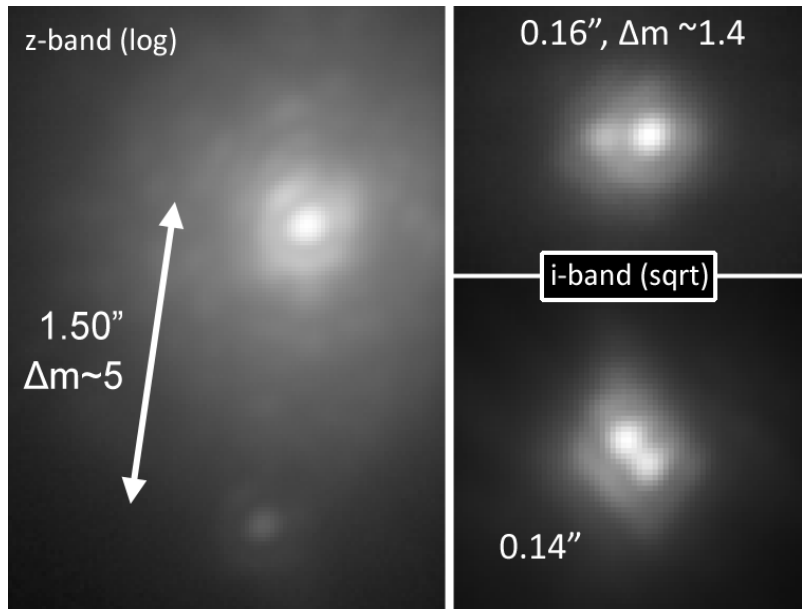

Figure 1. Images of binary stars captured with the prototype Palomar Robo-AO system.

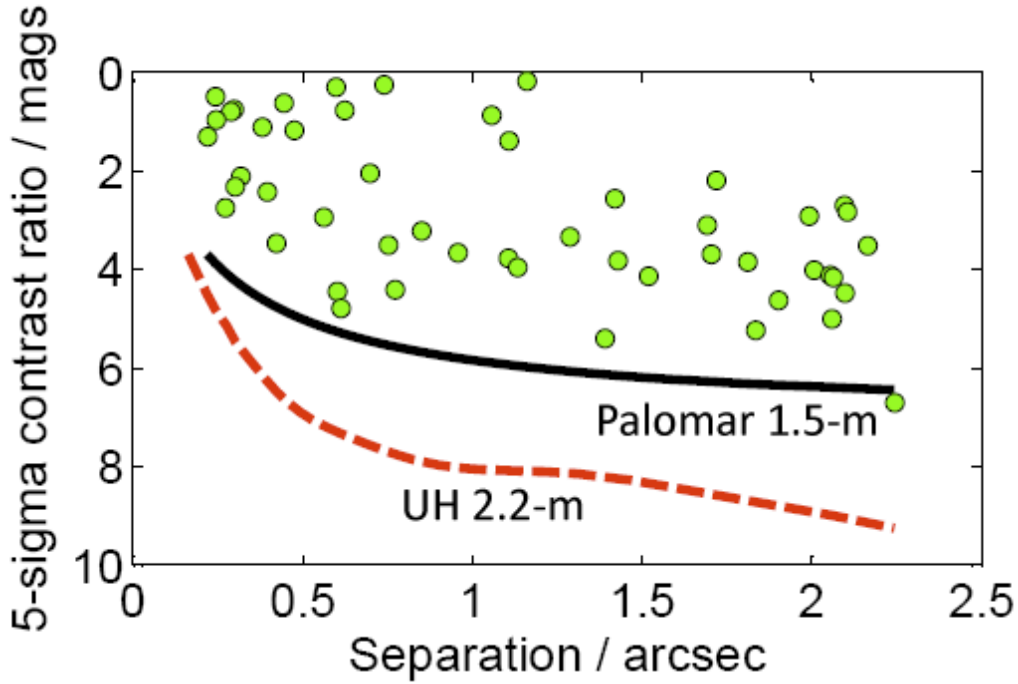

Figure 2. Measured visible 5-σ contrast ratio vs. separation for the prototype Robo-AO system in the best 25% conditions (black) and the predicted contrast achievable with the UH 2.2-m Robo-AO system in median, 50%, conditions (dashed red). The green circles represent actual companions detected as part of our survey of 715 candidate exoplanet host stars[6].

Sky coverage for imaging objects that are too faint to be used as tip-tilt guide sources will be modest (although the new Robo-AOs will be able to guide on stars a magnitude fainter than the prototype due to the larger apertures). At a galactic latitude of 30°, by guiding on a $m_V$=17 star in the visible, 16% and 9% of the sky can be accessed with the visible and infrared cameras, respectively, with up to an additional 90 and 80 milli-arcsec (mas) of RMS tip-tilt (two-axis) error to be added in quadrature. Guiding on a $m_V$=17 (MV) star in the infrared will allow access to 9% and 5% of the sky on the visible and infrared cameras, at the cost of up to 80 and 60 mas of RMS tip-tilt error respectively.

## 3. UPGRADED HARDWARE

### 3.1 Design

The design of the upgraded Robo-AO systems is based on the successful prototype[1,24] and includes several improvements to the performance and functionality. The systems will comprise a laser projector, a Cassegrain mounted adaptive optics system with imaging cameras and external instrument ports, and a set of computers and electronics.

### 3.2 Laser guide star

The laser projector will be a copy of the prototype laser projector which has a compact, 1.5 m × 0.4 m × 0.25 m, relatively lightweight, ~70kg, enclosure. The projector enclosure will attach to the side of the host telescope using a custom adapter bracket. Inside the projector are a commercial pulsed 12-W ultraviolet laser (λ=355nm) laser; a redundant shutter for safety; and an uplink tip-tilt mirror to both stabilize the apparent laser beam position on sky and to correct for up to 2' of differential telescope flexure. A bi-convex lens on an adjustable focus stage expands the laser beam to fill a 15 cm output aperture lens, which is optically conjugate to the tip-tilt mirror. The output lens focuses the laser light to a 9 km line-of-sight distance (and we will be using a nominal 400-m range-gate).

Because the UV laser is invisible to the naked eye, it has been approved for use without human spotters by the Federal Aviation Authority; however, coordination with U.S. Strategic Command is still necessary to avoid illuminating critical space assets. Instead of clearance for individual targets, we have implemented a strategy whereby we request open times for azimuth and elevation ranges (each 2.5°×2.5°) over the entire sky above 40° elevation, ensuring that at any given time there are targets available to observe with the laser. This also allows us to observe new targets, e.g., transients, SNe, on the fly.

Two improvements to the laser system will be implemented. A periscope will be added to the end of the telescope tube to jog the beam on-axis. In addition to reducing measurement error by minimizing perspective elongation of the returning laser light, this will prevent the laser beam from impinging on the equatorial-telescope domes. While the laser projector pointing is stable to the level correctable by the uplink tip-tilt mirror, the periscope will include active pointing to compensate for deterministic mechanical flexure. We will also implement passive and/or active controls on the laser projector focus to compensate for the temperature dependent focus drift found with the prototype (e.g. using an internal invar or carbon fiber, as opposed to aluminum, honeycomb breadboard).

### 3.3 Cassegrain mounted adaptive optics system

The adaptive optics system and science cameras will reside within a Cassegrain mounted structure of approximate dimensions 1 m × 1 m × 0.2 m, and will be conceptually similar to the layout of the prototype system[24]. Light from the telescope will enter the instrument and be intercepted by a fold mirror which directs the field to a dual off-axis parabolic (OAP) mirror relay. The first fold mirror will be on a linear motorized stage that can be moved out of the way to enable a pass-through mode for seeing-limited instruments to be co-mounted with the adaptive optics system. For the UH 2.2-m, this will obviate the need to ever remove the adaptive optics system and is made possible by the 2.2-m telescope nominal 313 mm back focal distance. Also with the fold mirror moved, a calibration source that matches the host telescope focal ratio and exit pupil position will simultaneously mimic the ultraviolet laser focus at 9 km and a blackbody source at infinity.

The first OAP will image the telescope pupil onto the deformable mirror. After reflection off of the deformable mirror, the UV laser light will be selected off with a dichroic mirror and sent to a UV optimized wavefront sensor. The visible and infrared light will pass through the dichroic and will be refocused by another OAP. The light will then be relayed by a second OAP relay which includes a tip-tilt corrector and an atmospheric dispersion corrector (for 450nm $< \lambda < 1.8\mu$m or 2.3 μm). The final relay element will create a telecentric F/41 beam which will be split by a dichroic mirror at $\lambda = 950$ nm. Each channel will have an imaging camera which can double as a tip-tilt sensor as well as a fold mirror on a linear stage which can fold the beam out of the instrument to ports where external instruments can be mounted.

### 3.4 Upgraded deformable mirror

We will upgrade the deformable mirror from a Boston Micromachines 12×12 actuator continuous face-sheet mirror with 3.5 μm of stroke to an IrisAO 489 actuator segmented piston-tip-tilt (PTT) mirror (13 hexagonal segments across) with 5 μm of stroke. While this nominally only increases the natural fitting of the wavefront from 12 to 14 actuators across the pupil, the coefficient of fitting error reduces from 0.28 for a continuous mirror to ~0.18 for a PTT mirror [25], and is therefore equivalent to a 17 actuator across continuous mirror. This larger actuator count deformable mirror is necessary to account for the larger telescope apertures. Another benefit of the IrisAO mirror is the ability to apply custom dielectric coatings instead of the bare aluminum or gold available from Boston Micromachines. The UV silver coating from Spectrum Thin Films (>98% R from 320nm<$\lambda$<2500nm) we applied to our prototype first fold and first OAP will be applied to the IrisAO mirror, increasing reflectivity by at least 10%.

### 3.5 Upgraded UV optimized wavefront sensor

The prototype Robo-AO system uses an 11×11 rectilinear Shack-Hartmann wavefront sensor. Range gating of the pulsed laser, to block Rayleigh scattered light above and below the focused beam waist, is accomplished with a Pockels cell between two crossed linear polarizers. The detector is a UV optimized CCD39 (80×80 pixels; 72% quantum efficiency at 355 nm). We bin the pixels by a factor of 3 (binned read noise of 6e$^-$ at 1.2kHz) and use 2×2 binned pixels to calculate the slope of each subaperture.

For the new Robo-AO systems, the 13 deformable mirror segments will be matched with a 13-across hexagonal lenslet array at a pupil. To sample each lenslet at 2×2 pixels on a detector requires a minimum of 30×28 pixels for the hexagonal geometry. We will use a cylindrical lens immediately after the lenslet array to address the 1:1.15 packing ratio of hexagonal lenslets, with negligible loss of image quality.

We will take advantage of CCD developments at MIT/Lincoln Labs, sponsored by Starfire Optical Range, for an upgraded wavefront sensor detector. MIT/LL will be providing a UV optimized (>50% QE) CCID82 device which is a version of their CCID75 device[26,27] built for adaptive optics applications (160×160 format, with 3e$^-$ of read noise at a 1.3kHz full frame rate) with additional processing to add an electronic shutter (an improved shutter to that used with the CCID18 at the MMT[28]). In addition to being a superior detector, and eliminating the need to have an external Pockels cell shutter, demonstrating the CCID82 on-sky will be crucial for other large telescope adaptive optics systems planning to use Rayleigh laser guide stars, e.g. extending the faintness limit of extreme AO systems (e.g. PULSE upgrade to PALM-3000; ref. 29) and for ground-layer adaptive optics systems[30].

### 3.6 Science cameras / tip-tilt sensors

The new Robo-AO systems will take advantage of recent technological advancements in visible and infrared cameras which can be used for imaging and tip-tilt sensing. For example, Nüvü Camera's HNü 1024 camera and Andor Technology's iXon Ultra 888 camera use the same E2V CCD201 electron-multiplying CCD as the prototype Robo-AO Andor 888 camera but have faster full-frame readout rates at 17.5 and 25.6 Hz respectively – this provides better post-facto image motion correction on brighter targets.

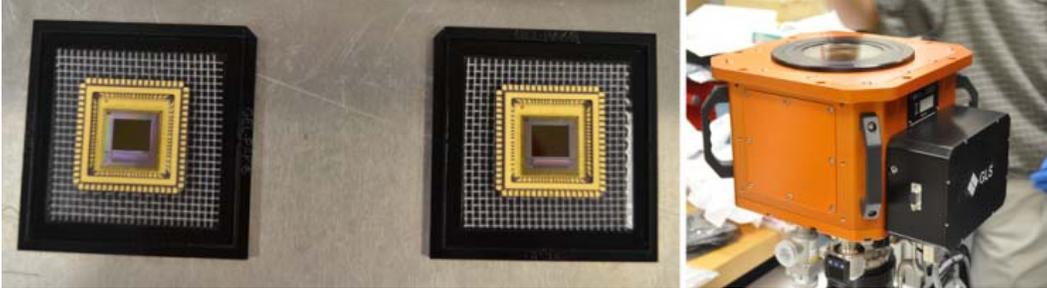

Figure 3. Left: the two SAPHIRA detectors at UH. Right: compact cryostat for on-sky detector testing.

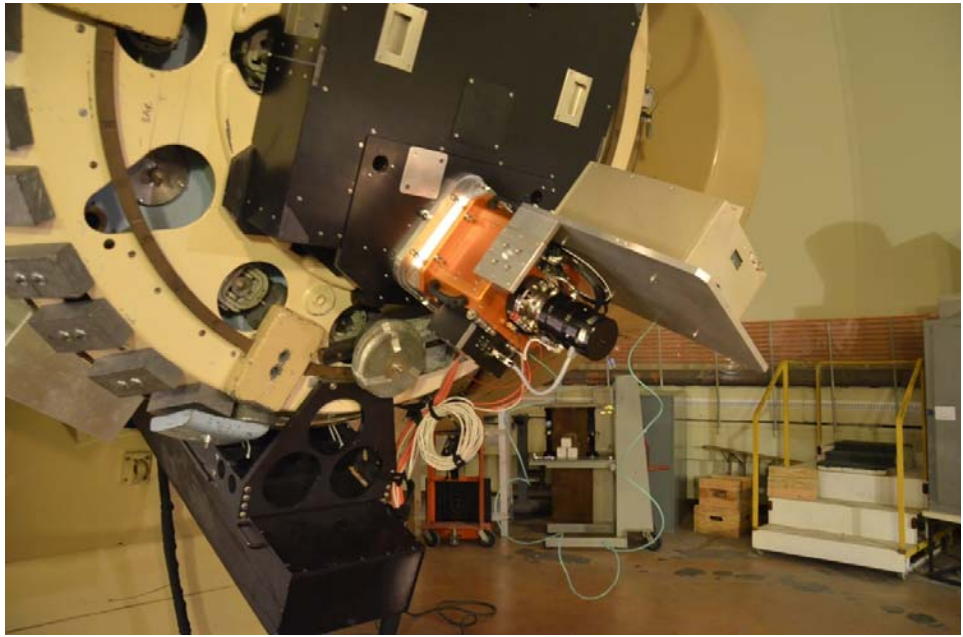

Figure 4. The SAPHIRA detector installed in a cryostat (orange) mounted to the external infrared port of the prototype Robo-AO system during initial testing in June 2014.

For the infrared camera, we will be using a Selex Avalanche PHotodiode InfraRed Array (SAPHIRA) that was originally developed by ESO and Selex for the GRAVITY instrument on the VLTI[31] and have been adopted by ESO for future AO wavefront sensing applications. Don Hall has been funded by the NSF to further develop these 256×256 devices for infrared tip-tilt sensing (AST-1106391) and by NASA for potential space applications (NNX13AC13G) as they are capable of very fast readouts, with 32 parallel outputs, and very low noise, <3e-, via avalanche amplification of the signal in the HgCdTe before being read out. Two devices have already been delivered (Fig. 3; left) and two more devices are being manufactured with improved processing as part of an ongoing development program. We have tested the devices in hand with Leach electronics in a dark dewar and in a compact cryostat (Fig. 3; right) with a cold, fixed, thermally-blocking H-band filter. The camera system was tested initially in a speckle shift-and-add imaging mode at the IRTF's native F/37.2 focus in April 2014, during which we demonstrated diffraction-limited H-band imaging performance[13]. In June 2014, we brought the infrared camera system to the prototype Robo-AO system (Fig. 4) to demonstrate both infrared imaging[13] and tip-tilt sensing as part of the robotic science operations. During the summer of 2014, we will specifically use the infrared tip-tilt capability to observe hundreds of the cool and faint ($m_V > 16$) *Kepler* candidate exoplanet hosts at visible wavelengths as part of our ongoing surveys[6].

In each of the visible and infrared channels of the new Robo-AO systems, there will be fold mirrors on motorized stages which can be used to fold the corrected beam out of the instrument to ports which can mount external instruments. For the UH Robo-AO, it should be noted that we can use instruments that already exist, such as the visible OPTIC camera with two 2k×4k orthogonal transfer CCDs (0.035"/pixels at F/41, 143" square field of view) or future instruments such as our new STA1600 CCD camera (0.020"/pixel, 3.5' FoV; funded by NASA) or our infrared H4RG-15 test camera (a 4k device with similarly sized pixels to OPTIC) being developed by Don Hall at UH (AST-0804651). An eyepiece can also be mounted in the visible port to support outreach and educational activities as previously done with the prototype[32].

### 3.7 Robotic software

The new Robo-AO system will reuse the robotic software developed for the prototype[33,34]. A single computer commands the AO system, the laser guide star, visible and near-infrared science cameras, the telescope, and other instrument functions. The easy-to-modify C++ software was designed from the outset to be very modular: the software to control each hardware subsystem was developed as a set of individual modules; and small standalone test programs have been created to test each of the hardware interfaces. This modular design allows the individual subsystems to be stacked together into larger modules, which can then be managed by other facets of the robotic control system.

The robotic control system executes tasks that have previously been performed manually; e.g., to start an observation, the central robotic control daemon will point the telescope, move the science filter wheel, and configure the science camera, laser, and AO system, during the telescope slew. After settling, an automated laser acquisition process executes, and an observation begins. A redundant safety system manages laser propagation onto the sky, and stops laser operations if any errors occur. After completion of an observation the control system will query the intelligent queue for the next target to observe. The intelligent queue is able to pick from targets in a directory structure organized by scientific program, with observation parameters defined as .xml files. All targets from all programs are initially considered "available", and the queue uses an optimization routine based on priority, slew time and cutoffs to determine the next target to observe. An analysis of the logs over the past year shows that the instrument takes 40-42 seconds to set up the instrument for each observation (not including an average of 40 s of slew time).

To have the new Robo-AO systems work as the prototype does now, we will write new modules for the upgraded pieces of hardware: the visible, infrared and wavefront sensor cameras; the deformable mirror; and the UH 2.2-m and IRTF telescope control systems. An existing fast-frame rate visible light camera data reduction daemon currently reduces all data shortly after it is acquired[1]. We will create new automatic data reduction daemons to reduce data from the infrared camera, as well as handle long-integration data from each of the cameras. We will be developing additional extensions for the intelligent queue so that it can poll the Maunakea weather station, seeing monitor and laser traffic control system[35] as part of its decision making process. We will be coordinating closely with the other observatories on the safe use of the UV laser via the Maunakea Laser Operators Group (which includes members from the visible and infrared observatories on Maunakea). We will also be developing a protocol for targets-of-opportunity to be remotely and automatically add into the queue at high priority, e.g. SNe discovered by ATLAS.

## 4. ADAPTIVE OPTICS PERFORMANCE

### 4.1 Robo-AO Error Budget

We have maintained detailed error budgets for the expected adaptive optics performance of the prototype Robo-AO system under different observing conditions and have validated the performance on sky[32]. This error budget was originally developed by Richard Dekany (Caltech) and collaborators, and has been additionally validated against on-sky performance of the Keck and 5-m Hale laser AO systems, as well as the prototype Robo-AO system. Using this same tool, we have estimated the performance of the UH Robo-AO system, Table 2, and IRTF Robo-AO system, Table 3. The error budgets use measured Maunakea $C_n^2(h)$ profiles derived from a combination of the Gemini ground-layer study[36] and an analysis, by us, of the first three years of data from the Maunakea summit MASS/DIMM seeing monitor running since fall 2009. A conservative 0.4" of additional dome seeing has been added to the $C_n^2(h)$ profiles.

## 4.2 UH 2.2-m Robo-AO

We expect dramatically increased imaging performance in all visible and infrared wavelengths: in median conditions, and pointing nearly at zenith, the UH Robo-AO will deliver an i-band Strehl ratio of 18% and an image width of 0.08" vs. 10% and 0.15" for the prototype Robo-AO. At the short wavelength range, wavelengths as short as 400 nm will receive significant image sharpening, opening up scientific opportunities where few adaptive optics systems have been able to operate. For guide sources brighter than $m_V$=17 (MV), near diffraction-limited-image-width performance is limited by high-order wavefront errors from the laser system. Beyond $m_V$=19.5 (MV), tip-tilt errors start to dominate, affecting the overall AO correction.

Table 2. Example error budget for the UH Robo-AO under different seeing conditions ($r_0$) and for different zenith angles (z) assuming an on-axis $m_V$=17 MV star for tip-tilt sensing and on-axis science target. Performance in the visible bands (light blue) uses an infrared tip-tilt signal with an open (Y+J+H) filter. Performance in the infrared bands (green) uses a visible tip-tilt signal with an r+i+z filter. Measurement error arises from finite laser photoreturn, WFS read noise, sky noise, dark current, and other factors. Focal anisoplanatism is an error arising from the finite altitude of the Rayleigh laser resulting in imperfect atmospheric sampling. Multispectral error arises from differential refraction of UV and visible/NIR rays. 'mas' indicates milli arc seconds, (") indicates arc seconds.

| Percentile Seeing | 25% | 50% | 50% | 75% |
|---|---|---|---|---|
| Atmospheric r0 | 22.1 cm | 16.8 cm | | 10.3 cm |
| Effective seeing at zenith (with dome seeing) | 0.69" | 0.80" | | 1.00" |
| Zenith angle | 15 degrees | 15 degrees | 45 degrees | 35 degrees |
| **High-order Errors** | \multicolumn{4}{c}{Wavefront Error (nm)} | | | |
| Atmospheric Fitting Error | 38 | 43 | 51 | 56 |
| Bandwidth Error | 36 | 42 | 49 | 54 |
| High-order Measurement Error | 24 | 27 | 30 | 34 |
| LGS Focal Anisoplanatism Error | 75 | 103 | 124 | 160 |
| Multispectral Error | 3 | 3 | 94 | 46 |
| Scintillation Error | 12 | 15 | 27 | 27 |
| WFS Scintillation Error | 10 | 10 | 10 | 10 |
| Uncorrectable Tel / AO / Instr Aberrations | 38 | 38 | 38 | 38 |
| Zero-Point Calibration Errors | 34 | 34 | 34 | 34 |
| Pupil Registration Errors | 21 | 21 | 21 | 21 |
| High-Order Aliasing Error | 13 | 14 | 17 | 19 |
| DM Stroke / Digitization Errors | 1 | 1 | 1 | 1 |
| **Total High Order Wavefront Error** | **112 nm** | **136 nm** | **185 nm** | **198 nm** |
| **Tip-Tilt Errors** | \multicolumn{4}{c}{Angular Error (mas)} | | | |
| Tilt Measurement Error | 11 | 11 | 13 | 14 |
| Tilt Bandwidth Error | 8 | 11 | 9 | 11 |
| Science Instrument Mechanical Drift | 6 | 6 | 6 | 6 |
| Residual Telescope Pointing Jitter | 2 | 2 | 2 | 2 |
| Residual Centroid Anisoplanatism | 1 | 1 | 2 | 2 |
| Residual Atmospheric Dispersion | 1 | 1 | 4 | 3 |
| **Total Tip/Tilt Error (one-axis)** | **15 mas** | **16 mas** | **17 mas** | **19 mas** |
| **Total Effective Wavefront Error (IRTT)** | 129 nm | 155 nm | 195 nm | 208 nm |
| **Total Effective WFE (VISTT)** | 124 nm | 157 nm | 195 nm | 209 nm |

| Spectral Band | λ | λ/D | Strehl | FWHM | Strehl | FWHM | Strehl | FWHM | Strehl | FWHM |
|---|---|---|---|---|---|---|---|---|---|---|
| g' | 0.47 μ | 0.044" | 6% | 0.06" | 2% | 0.07" | 1% | 0.11" | 0% | 0.49" |
| r' | 0.62 μ | 0.058" | 18% | 0.07" | 9% | 0.07" | 6% | 0.08" | 1% | 0.13" |
| i' | 0.75 μ | 0.070" | 30% | 0.08" | 18% | 0.08" | 14% | 0.08" | 5% | 0.10" |
| J | 1.25 μ | 0.117" | 64% | 0.12" | 54% | 0.12" | 45% | 0.13" | 33% | 0.13" |
| H | 1.64 μ | 0.153" | 76% | 0.16" | 69% | 0.16" | 61% | 0.16" | 51% | 0.16" |

## 4.3 IRTF Robo-AO

Imaging performance of the IRTF Robo-AO system will be similar in delivered wavefront error to that of the prototype Robo-AO system at Palomar[32], mainly due to the much increased laser focal anisoplanatism error. Where the IRTF Robo-AO system excels is with system the faintness limit being extended by almost 1.5 magnitudes, and the angular resolution which is twice as sharp under comparable percentile seeing conditions. For guide sources brighter than $m_V$=18 (MV), near diffraction-limited-image-width performance is limited by high-order wavefront errors from the laser system. Beyond $m_V$=20 (MV), tip-tilt errors start to dominate, affecting the overall AO correction.

Table 3. Example error budget for the IRTF Robo-AO under different seeing conditions ($r_0$) and for different zenith angles (z) assuming an on-axis $m_V$=18 MV star for tip-tilt sensing and on-axis science target. Performance in the visible bands (i.e. r' and i'-bands) uses an infrared tip-tilt signal with an open (Y+J+H) filter. Performance in the infrared bands (i.e.. H-band) uses a visible tip-tilt signal with an r+i+z filter. Similar to Table 2, the total tip-tilt errors are effectively the same between the infrared and visible when guiding on an MV type star (the increasing sky brightness in the near-infrared is compensated by the increasing brightness of the tip-tilt star in the corresponding wavebands).

| | Percentile Seeing | 25% | 50% | 50% | 75% |
|---|---|---|---|---|---|
| | Atmospheric r0 | 22.1 cm | 16.8 cm | 16.8 cm | 10.3 cm |
| Effective seeing at zenith (with dome seeing) | | 0.69" | 0.80" | 0.80" | 1.00" |
| | Zenith angle | 15 degrees | 10 degrees | 40 degrees | 25 degrees |
| **High-order Errors** | | \multicolumn{4}{c|}{Wavefront Error (nm)} | | | |
| Atmospheric Fitting Error | | 49 | 56 | 63 | 69 |
| Bandwidth Error | | 36 | 42 | 47 | 51 |
| High-order Measurement Error | | 26 | 29 | 30 | 35 |
| LGS Focal Anisoplanatism Error | | 96 | 130 | 152 | 192 |
| Multispectral Error | | 5 | 3 | 32 | 16 |
| Scintillation Error | | 12 | 15 | 23 | 23 |
| WFS Scintillation Error | | 10 | 10 | 10 | 10 |
| Uncorrectable Tel / AO / Instr Aberrations | | 38 | 38 | 38 | 38 |
| Zero-Point Calibration Errors | | 34 | 34 | 34 | 34 |
| Pupil Registration Errors | | 21 | 21 | 21 | 21 |
| High-Order Aliasing Error | | 16 | 19 | 21 | 23 |
| DM Stroke / Digitization Errors | | 2 | 2 | 2 | 2 |
| **Total High Order Wavefront Error** | | **131 nm** | **162 nm** | **197 nm** | **223 nm** |
| **Tip-Tilt Errors** | | \multicolumn{4}{c|}{Angular Error (mas)} | | | |
| Tilt Measurement Error | | 10 | 12 | 12 | 14 |
| Tilt Bandwidth Error | | 7 | 8 | 9 | 10 |
| Science Instrument Mechanical Drift | | 6 | 6 | 6 | 6 |
| Residual Telescope Pointing Jitter | | 2 | 3 | 2 | 2 |
| Residual Centroid Anisoplanatism | | 1 | 2 | 2 | 2 |
| Residual Atmospheric Dispersion | | 0 | 0 | 2 | 1 |
| **Total Tip/Tilt Error (one-axis)** | | **14 mas** | **15 mas** | **16 mas** | **18 mas** |
| **Total Effective Wavefront Error** | | **153 nm** | **182 nm** | **205 nm** | **241 nm** |

| Spectral Band | λ | λ/D | Strehl | FWHM | Strehl | FWHM | Strehl | FWHM | Strehl | FWHM |
|---|---|---|---|---|---|---|---|---|---|---|
| r' | 0.62 μ | 0.04" | 10% | 0.06" | 3% | 0.06" | 1% | 0.08" | 0% | 0.36" |
| i' | 0.75 μ | 0.05" | 19% | 0.06" | 10% | 0.06" | 5% | 0.07" | 2% | 0.10" |
| H | 1.64 μ | 0.11" | 69% | 0.12" | 59% | 0.12" | 34% | 0.12" | 41% | 0.12" |

# 5. DISCUSSION

## 5.1 Clone of Robo-AO

A clone of the current Robo-AO system is currently being developed for the 2-m IUCAA Girawali Observatory telescope[14] in Maharashtra, India. Adaptations in optical prescriptions have been made to accommodate the slightly different F/#, F/10 vs. F/8.75, different mounting interface and telescope interface software. No other major changes are anticipated - minimizing further development and control costs. The system is expected to see first light in 2015.

## 5.2 Natural-guide star version of Robo-AO

The KAPAO (KAPAO: A Pomona Adaptive Optics) project is an ongoing instrumentation effort between Pomona College, Sonoma State University, Caltech and Harvey Mudd College, to develop and deploy a low-cost, remote-access, natural guide star adaptive optics system for the Pomona College Table Mountain Observatory (TMO) 1-m telescope[15,16]. The system uses the core Robo-AO adaptive optics control software and was primarily built using undergraduate labor, resulting in several senior theses. The system offers simultaneous dual-band, diffraction-limited imaging at visible and near-infrared wavelengths and will delivers an order of magnitude improvement in point source sensitivity and angular resolution relative to the current seeing limit.

The primary scientific objectives for developing AO capabilities on the 1-meter telescope are: (1) To enhance ongoing TMO science programs with a factor of 5-10 improvement in image resolution and open up new avenues of research and new collaborations that will maximally benefit from high resolution monitoring capabilities. (2) To serve as an on-sky testbed for advancing new wavefront sensing technologies such as pyramid wavefront sensing , spatial filtered wavefront sensors and Fourier-based reconstructors.

## 5.3 All-sky Robo-AO network

The prototype Robo-AO at Palomar has been crucial in validating the current sample of KOIs, having observed over half of the almost 3,000 host stars (as of June 2014). A complementary follow-on mission to *Kepler* is the Transiting Exoplanet Survey Satellite (TESS; scheduled for launch in 2017) led by MIT[22]. TESS will execute a shallower survey compared to *Kepler*, with the majority of objects $m_V < 16$, but over the entire sky, surveying the northern sky in year 1 and the southern sky in year 2. It is estimated that there may be as many as ten or more times as many transit signals discovered by TESS during its mission lifetime – which could all be validated by Robo-AO, in an analogous way to the KOIs, in less than a year. To make this happen, a North-South network of facility-class second generation Robo-AO systems would be necessary: in addition to the 2.2-m UH and 3-m IRTF telescopes on Maunakea, we are looking for partners for a facility-class Robo-AO in the southern hemisphere, preferably on a robotic 1.5 to 3 m sized telescope.

# ACKNOWLEDGEMENTS

The prototype Robo-AO system was developed by collaborating partner institutions, the California Institute of Technology and the Inter-University Centre for Astronomy and Astrophysics, and with the support of the National Science Foundation under Grant Nos. AST-0906060 and AST-0960343, the Mt. Cuba Astronomical Foundation and by a gift from Samuel Oschin. Ongoing science operation support of the prototype Robo-AO system is provided by the California Institute of Technology, the University of Hawai'i and the University of North Carolina at Chapel Hill. The wide-field infrared camera upgrade is supported by the California Institute of Technology and the Inter-University Centre for Astronomy and Astrophysics with funding from the National Science Foundation under Grant No. AST-1207891 and by the Office of Naval Research under grant N00014-11-1-0903. The high-speed infrared tip-tilt camera is supported by the University of Hawai'i and the National Science Foundation under Grant No. AST-1106391. C.B. acknowledges support from the Alfred P. Sloan Foundation. We thank Richard Dekany for use of the Wavefront Error Budget Tool. We are grateful to the Palomar Observatory staff for their extraordinary support of Robo-AO on the 1.5-m telescope, particularly S. Kunsman, M. Doyle, J. Henning, R. Walters, G. Van Idsinga, B. Baker, K. Dunscombe and D. Roderick.